\documentclass[fleqn,10pt,twocolumn]{wlscirep}

\usepackage{graphicx}
\usepackage{float}
\usepackage[squaren]{SIunits}
\usepackage{hyphenat}
\usepackage[bookmarks=false] {hyperref}
\usepackage{color}
\usepackage[capitalise]{cleveref} 
\crefname{section}{Sec.}{Secs.}
\Crefname{section}{Section}{Sections}
\usepackage{makecell}
\usepackage{physics}
\usepackage{color}
\usepackage{comment}

\newcolumntype{L}[1]{>{\raggedright\let\newline\\\arraybackslash\hspace{0pt}}p{#1}}
\newcolumntype{C}[1]{>{\centering\let\newline\\\arraybackslash\hspace{0pt}}p{#1}}
\newcolumntype{R}[1]{>{\raggedleft\let\newline\\\arraybackslash\hspace{0pt}}p{#1}}

\definecolor{ss_color}{rgb}{1,0,0}

\definecolor{tj_color}{rgb}{0,0,1}

\title{Observing quantum coherence from photons scattered in free-space}

\author[1,2,*]{Shihan Sajeed}
\author[1,2,$\dagger$]{Thomas~Jennewein}

\affil [1] {Institute for Quantum Computing, University of Waterloo, Waterloo, ON, N2L~3G1 Canada}
\affil [2] {Department of Physics and Astronomy, University of Waterloo, Waterloo, ON, N2L~3G1 Canada}
\affil[*] {shihan.sajeed@uwaterloo.ca}
\affil[$\dagger$]{thomas.jennewein@uwaterloo.ca}

\date{\today}

\begin{abstract}
Quantum channels in free-space, an essential prerequisite for  fundamental tests of quantum mechanics and quantum technologies in open space, have so far been based on direct line-of-sight because the predominant approaches for photon-encoding,  including  polarization and spatial modes, are not compatible with randomly scattered photons.   Here we demonstrate a novel approach to transfer and recover quantum coherence from scattered, non-line-of-sight photons analyzed in a multimode and imaging interferometer  for time-bins, combined with photon detection based on a 8x8 single-photon-detector-array.  The observed time-bin visibility for scattered photons remained at a high $95\%$ over a wide scattering angle range of $-45\degree$ to $+45\degree$, while the individual pixels in the detector array  resolve or track an image in its field of view of ca. 0.5 degrees. Using our method we demonstrate the viability of two  novel applications. Firstly, using scattered photons as an indirect channel for quantum communication thereby enabling non-line-of-sight quantum communication with background suppression, and secondly, using the combined arrival time and quantum coherence to enhance the contrast of low-light imaging and laser ranging under high background light.  We believe our method will instigate new lines for  research and development on applying photon coherence from scattered signals to quantum sensing, imaging, and communication in free-space environments.
\end{abstract}

\begin{document}
	\flushbottom
	\maketitle
	
	\thispagestyle{fancy}
	
\section*{Introduction}
\label{sec:intro}
Quantum coherence is a key ingredient in many fundamental tests and applications of quantum mechanics including quantum communication~\cite{rubenok2013}, characterization of single-photon sources~\cite{michler2000}, generation of non-classical states \cite{lee2002}, quantum metrology \cite{fulvio2015}, quantum teleportation~\cite{pirandola2015},  quantum fingerprinting~\cite{xu2015},  quantum cloning~\cite{irvine2004}, demonstrating quantum optical phenomena~\cite{pan2012}, and quantum computing~\cite{spring2013} etc. The ability to transfer quantum coherence via scattering surfaces and its successful recovery from scattered photons enhances several applications of quantum technologies. For instance, quantum communication capable of operating over a scattering channel could accommodate free space communication with non-line-of-sight between multiple users such as indoors around corners, or with short range links with moving systems. Furthermore, the photon coherence recovered from scattered  light could be utilized to improve noise performance in low-light and 3D imaging, around-the-corner imaging~\cite{gariepy2016}, velocity measurement~\cite{erksine1995}, light detection and ranging (LIDAR), surface characterization,  or biomedical sample identification.

Currently,  the predominant photon encoding used on free-space quantum channels is polarization, because it is not impacted by  turbulent atmosphere for clear line-of-sight transmission\cite{Hohn:1969cr}. When photons are scattered, however, their polarization states are inherently disturbed and the quantum encoding is degraded. A previous study~\cite{bourgoin2014} showed that the observed polarization visibility depends on the scattering-surface material, and even the best material (cinematic silver screen) showed a strong dependence on the photon scattering angles with only a total angle of less than $45\degree$ was suitable for quantum communications. Another approach to encode free-space channels is the use of higher-order spatial modes, recently utilized for intracity quantum key distribution \cite{sit2017}, yet these photon states are directly impacted by wavefront distortion and are expected to completely vanish upon random scattering from a surface~\cite{patterson2005,malik2012}.

Here we utilize quantum coherence encoded in time-bins~\cite{brendel1999,stucki2005,rubenok2013} which -- although widely used for single-mode optical fibers~\cite{townsend1993,muller1997,inoue2002,stucki2005,stucki2009} -- has rarely  been demonstrated for free-space channels until recently~\cite{jin2018,jin2019,vallone2016}. This encoding is robust upon scattering. The multimode states of light have been utilized using field-widened interferometers, or imaging interferometers, thereby solving the wavefront distortions caused by turbulent media. Such interferometers have the additional benefit of preserving an image. (An alternate approach could be to convert the multimode optical beam back into a single-mode at the receiver by means of adaptive optics, however at the cost of additional losses~\cite{cao2020}). We thus implemented an imaging time-bin interferometer equipped with a single-photon-detector-array (SPDA) sensor, with 8x8 pixels covering a field of about $0.5\degree$. The photon detector achieves high temporal precision of $\approx 120~\pico\second$, combined with the ability to spatially resolve the field-of-view with excellent time-bin visibility across the whole sensor area. We demonstrate that our system allows imaging a target that was illuminated with photons prepared with specific phase-signatures, that are recovered from the scattered photons with excellent phase coherence by the illuminated pixels. We discuss the viability of our method in the context of two relevant applications. 

\section*{Experiment}
\begin{figure}[tb]
	\centering
	\includegraphics[width= \columnwidth]{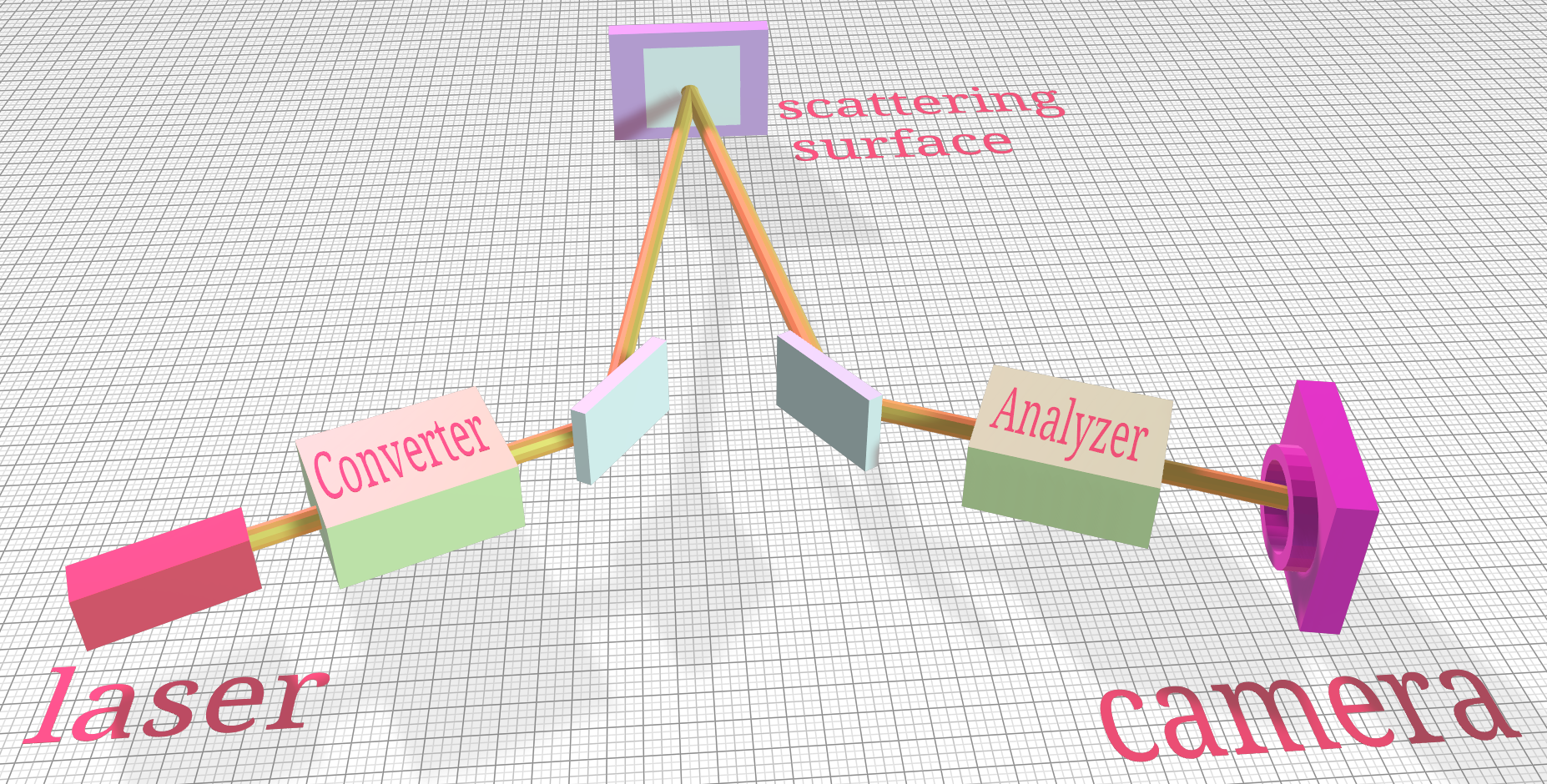}
	\caption{The experimental setup. Optical pulses from a laser are sent through a phase Converter, which creates the initial time-bin states, while the multi-mode Analyzer measures the signals scattered off the target (regular white paper). A single-photon-detector-array is used as the detection device, with 8 x 8 individual pixels which are time-tagged separately. During the initial alignment, the incident angle  = reflected angle  = $25~\degree$.}
	\label{fig:setup_time_bin} 
\end{figure}

\begin{figure}[htp!]
	\centering
	\includegraphics[width = 0.95\columnwidth]{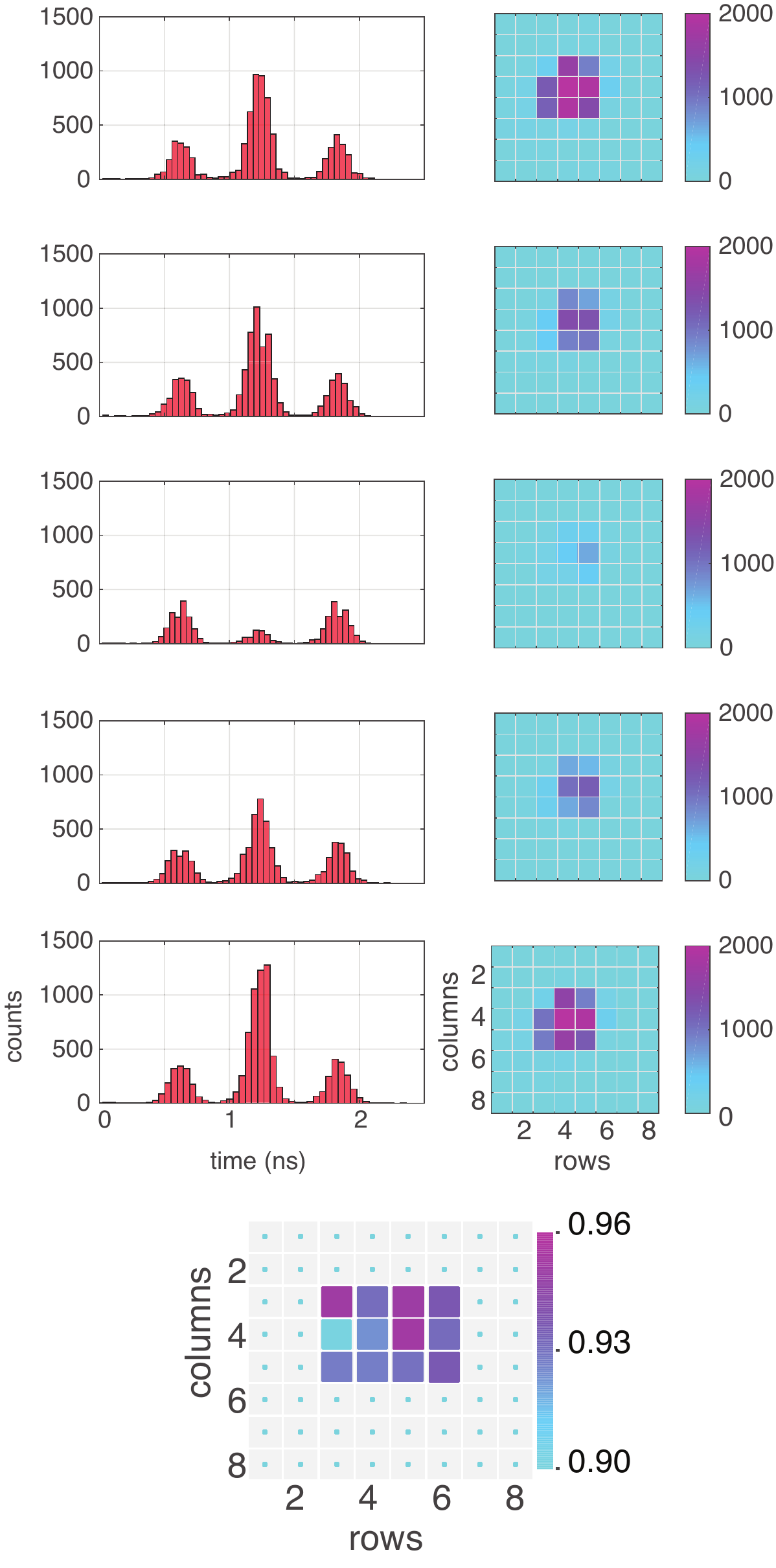}
	\caption{Observed interference. Top left: Histograms of the arrival times of photons at pixel number $(4,4)$ (row 4, column 4). The three separate peaks correspond to photons coming via SS (right), SL or LS (middle), and LL (left) paths (see text for more detail). Five different phase-instances are shown. The visibility at this pixel -- calculated by curve-fitting -- was $V_{(4,4)} \approx 0.95$. Top right: Middle-peak intensity at the illuminated pixels at the corresponding phase-instances. Bottom: visibilities of the illuminated pixels calculated after fitting the curves. The visibilities range from $0.9-0.95$. Both the color and the size of the \emph{sqaure} marker in each pixel area are indicators of the visibility. Pixel number $(1,1)$ was used as a trigger. The temporal precision was $\approx 120~\pico\second$.}
	\label{fig:spatial_temporal}
\end{figure}

The experimental setup is shown in \cref{fig:setup_time_bin} (see Methods for additional details). Each pulse from the laser passes through the Converter -- an unbalanced Michelson interferometer (UMI) -- that converts it into two coherent pulses separated by a time delay according to the path difference $\Delta_C$. These signals are sent towards the target sample covered with a diffusive material (regular white paper) acting as the \emph{scattering surface}, which can be rotated to vary the angle of incidence. Some of the photons scattered from the surface are captured and guided through the Analyzer, a second UMI with path difference $\Delta_A$. Finally, the photons emerging from the Analyzer are focused into a single-photon-detector-array (SPDA) containing $64$ single-photon avalanche photo-diodes -- hereafter referred to as pixels -- arranged in a $8\times 8$ row-column configuration, which are free-running and individually time-tagged. Each detected photon could have traversed one of the four possible paths: short-short (SS), short-long (SL), long-short (LS) and long-long (LL) with first (second) letter denoting the path taken inside Converter (Analyzer). The path-differences of the two UMI are tightly matched, i.e., $\Delta_C \approx \Delta_A$, and interference can be observed, as shown in \cref{fig:spatial_temporal}. Piezoelectric actuators are placed at the short arm of each interferometer to vary their respective phase. To compensate for variable angle of incidence and mode-distortion, a $118\milli\meter$ long glass cube with refractive index $1.4525$ is placed in the long arm of both interferometers. 
\begin{figure*}[htp]
	\centering
	\includegraphics[width= 0.9\textwidth]{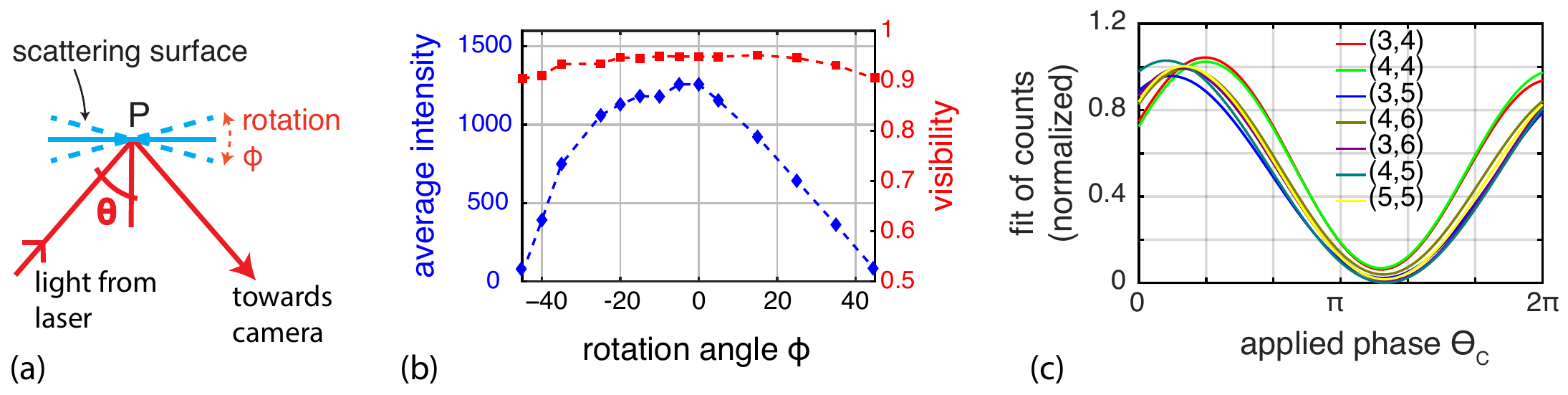}
	\caption{Visibility with scattered light and coherence among the pixels. a. The incident angle $\theta$ was varied by rotating the scattering surface about point $P$ by an angle $\phi$ from $-45\degree$ to $+45\degree$ while keeping the position of laser and camera constant. b. Variation of visibility and intensity with rotation angle. Here, rotation angle $\phi = 0$ corresponds to incidence angle of $\theta = 25\degree$ as shown in \cref{fig:setup_time_bin}. For rotation angles higher than $\pm 45\degree$, sufficient amount of photon could not be collected by the camera. The average intensity for each angle is shown on the right axis. The collection time for each data point is $20\sec$. c. Variation of counts versus phases applied at the Converter for the illuminated pixels. Pixel number $(1,1)$ was used as a trigger.}
	\label{fig:scattering_plot}
\end{figure*}

A sinusoidal phase difference was introduced between the two interfering (SL and LS) pulses by applying a $0.1~\hertz$ $10V$ peak-to-peak ramp voltage at the Analyzer piezoelectric actuator (i.e., the length of the $LS$ path was varied with respect to $SL$). The resultant outcomes are shown in \cref{fig:spatial_temporal}. The top left columns shows histograms of detection times -- with respect to trigger signal --  at one of the central pixels $(4,4)$ (row 4, column 4) for five different phase-instances. The three peaks correspond to photons coming via SS (right), SL or LS (middle) and the LL (left) paths.  The temporal precision was $\approx 120~\pico\second$. The measured visibility  -- calculated by fitting the curve -- was $V_{(4,4)} \approx 0.95$. The right column shows the corresponding middle-peak intensity for other illuminated pixels. In this case, only the events detected during the $0.6~\nano\second$ window -- centred around the middle-peak -- was post-selected. The data collection time was $1~\sec$ for the left and $0.1~\sec$ for the right columns. The visibilities of the illuminated pixels -- calculated after fitting the curves -- are shown at the bottom which range from $0.9-0.95$.



\subsection*{Robust and stable phase coherence under scattering angle variations}
	\label{robustness}
During the initial alignment, specular reflection was used by setting the incident angle = reflected angle = $25\degree$. We refer to the incident angle as  $\theta$. Then, using a rotational mount, the scattering surface was rotated about $P$ to vary $\theta$ while keeping the position of the camera fixed (see \cref{fig:scattering_plot}a). For different rotation angle $\phi$, the corresponding average intensity and visibilities are shown in \cref{fig:scattering_plot}b for pixel number $(4,4)$. For $\theta = 25\degree$ ($\phi = 0\degree$), the majority of photons collected by the camera was due to specular reflection. As $\theta (\phi)$ was varied, the intensity followed the typical scattering pattern consisting of specular and diffusive reflection \cite{tan2017}. At larger rotation angles $\phi$, the collected photons were mainly due to scattering from the diffused surface. For  $ \phi > \abs{45\degree}$, the amount of photons collected were too low. However, within the range  $\phi \in \{-45\degree, +45\degree\}$, although the number of detected photons varied, the visibility remained fairly constant at around $95\%$ with less than $10\%$ variation. The same behavour was observed for other illuminated pixels. 

\subsection*{Recovery of coherence while imaging}
\label{coherence}
As we will demonstrate, this approach offers the unique ability for individual pixels in this imaging analyzer to fully detect the coherence, while at the same time resolving spatial modes of the scattered photons. To ensure that the camera received mostly scattered photons, the scattering surface was rotated to set $\phi = 20~\degree$ (see \cref{fig:scattering_plot}a) at which the value of visibility was still close to $0.95$. Similar to \cref{fig:spatial_temporal}, only the pixels in the center of the sensor were illuminated. By applying a periodic voltage to the  piezoelectric actuator in the Converter interferometer, a sinusoidal phase variation was induced among the outgoing time-bin photons. For each phase $\theta_c$, the corresponding counts at the illuminated pixels were recorded. No voltage was applied at the Analyzer piezoelectric actuator. The fitted count-versus-phase curves for the illuminated pixels are shown in~\cref{fig:scattering_plot}c. The result verified that although each pixel saw a different set of spatial modes of the incoming light, the phase $\theta_c$ was preserved by the modes even after going through diffused scattering, and was successfully recovered from the scattered photons by each individual pixel independently.  The test was repeated by illuminating all  pixels of the SPDA  for different scattering angles $\phi$, yielding similar results. 


\subsection*{Quantum communication with scattered light}
\label{nlos_qc}
We now show the applicability of our setup in implementing quantum communication (QC) with scattered photons in a non-line-of-sight (nLOS) scenario. Here, nLOS refers to the fact that no direct path exists between sender and receiver; neither do they have access to any specular reflector, i.e. mirror surface, at any intermediate point. More specifically, we consider the implementation of the phase-encoded Bennett-Brassard 1984 (BB84) QKD protocol~\cite{bennett1984}.  The experimental setup was similar to \cref{fig:setup_time_bin} where the laser and Converter can be considered to be at sender's (Alice's) lab while the Analyzer and the camera at receiver's (Bob's) lab. Photons from Alice's laser were scattered off the diffused surface (white paper) before being measured by Bob. For our experiment, we rotated the scattering surface by $\phi  = 20~\degree$ to ensure that photons detected by Bob were mostly due to scattering from the diffusive surface. The four phase-encoded BB84 states sent by Alice are:
\begin{equation}
\ket{\psi_A} = \ket{e} + e^{i\theta_A} \ket{l}.
\end{equation}
Here, $\ket{e}~(\ket{l})$ denotes the state of the photon that took the early (late) path at the Converter, and $\theta_A \in \{0,\pi/2, \pi,3\pi/2\}$ is the phase difference applied at the Converter. These states can be generated by applying specific voltages to the piezoelectric actuator placed at the long arm of the Converter. Bob's measurement basis can be chosen by applying phase difference $\theta_B \in \{0, \pi\}$ at the Analyzer piezo actuator. \Cref{fig:scattering_plot}c shows the variation of counts versus $\theta_A$ at the illuminated pixels for Bob's measurement basis $\theta_B = 0$. Similar variation was observed when Bob's basis was changed to $\theta_B = \pi$ (not shown). The visibilities were in the range of $0.9-0.95$ for all the illuminated pixels. The result verifies that even though there was no direct line of sight between Alice and Bob, and the photons were being scattered by a diffused surface on their way from sender to receiver, they could still be used for implementing the BB84 protocol with high visibility. 
Our proof-of-principle implementation makes it an ideal candidate for real-world quantum communication applications over scattering and nLOS channels.

We emphasize that the SPDA sensor  enhances the robustness of the QKD receiver, in addition to allowing nLOS operation. Firstly, the imaging information for each pixel can be used for tracking a transmitter beam that feeds into an  active beam steering mechanism, or simply used to passively suppress background counts in processing. Secondly, the implementation of multiple photon detector pixels appears to be more robust against certain powerful quantum attacks. For example, the class of detector control attacks~\cite{lydersen2010a,lydersen2011} could be prevented by  adopting the method presented in \cite{gras2020}. The efficiency mismatch type attacks~\cite{sajeed2015a} -- where an eavesdropper attacks by modifying the spatial modes of the incoming light -- could be easily detected as changing the spatial mode will change the spatial distribution of detection events.

\subsection*{Low-light imaging in high photon background}
\begin{figure*}[t]
	\centering
	\includegraphics[width=\textwidth]{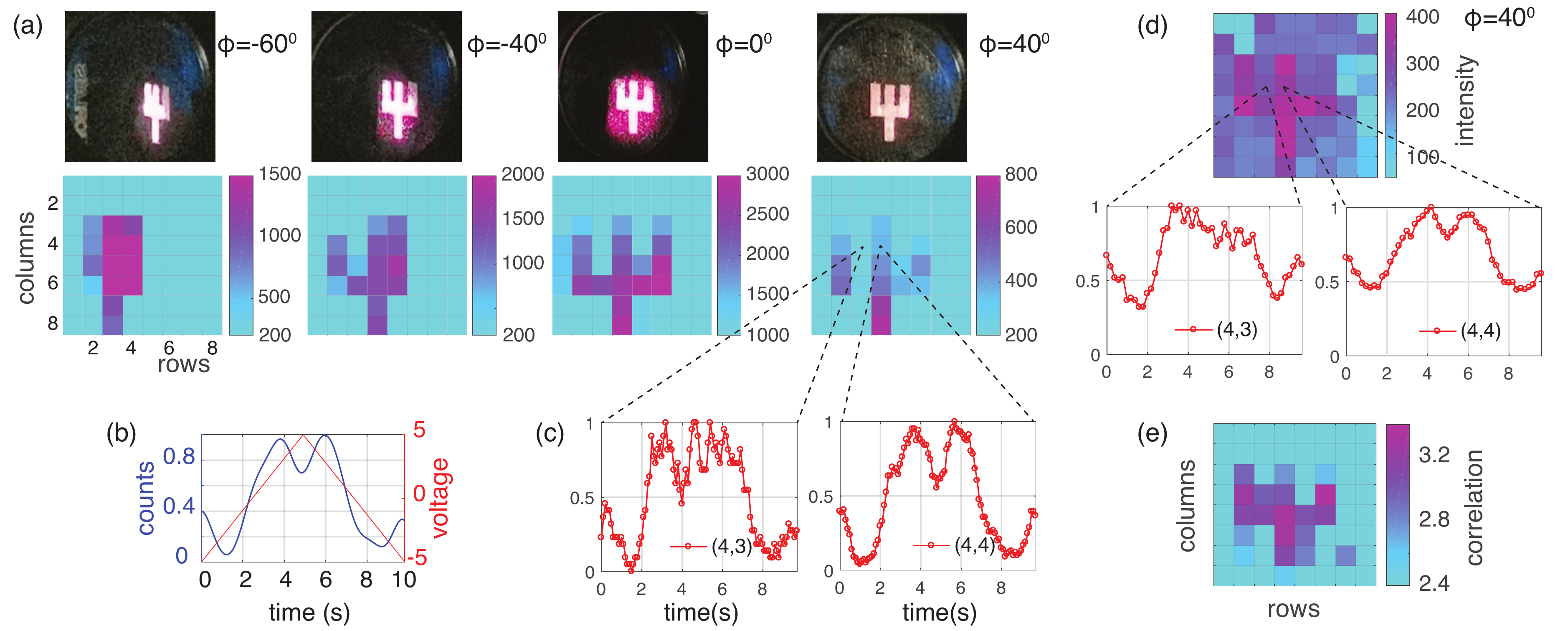}
	\caption{Enhancing the contrast of an image. a. The top and bottom row show the illuminated object (size $4 \times 3.6\milli\meter$) -- with the scattering surface rotated at four different angles $\phi \in \{ -60\degree,-40\degree,-60\degree,40\degree\}$ -- and the corresponding images observed in the camera. b. The variation of detector counts as a function of the applied phase-signature. This is the expected count pattern, the reference pattern, for high signal-to-noise ratio scenario. c. The observed pattern in response to the phase-signature for two neighbouring pixels $(4,3)$ and $(4,4)$ for rotation $\phi = 40\degree$. d. Image captured by the SPDA imager in the low-light and noisy environment. The observed pattern at the same two neighbouring pixels $(4,3)$ and $(4,4)$ are also shown. e. The image reconstructed by cross-correlating the observed pattern with the reference pattern (see text and Methods for more details).} 
	\label{fig:imaging_with_intensity}.
\end{figure*}
It is important to note  that the ability to reveal quantum coherence  can be directly used to enhance the contrast of an image under a low-light and noisy environment. The experimental setup is similar to \cref{fig:setup_time_bin}, but now the laser focusing condition is adjusted such that all  pixels of the SPDA were illuminated. \cref{fig:imaging_with_intensity} a shows the conventional intensity-based images of the object for four different rotations of the scattering surface. For these tests the scattering surface was rotated to $\phi = 40\degree$, ensuring that the specular reflection no longer reaches the camera, and the detected photons were only due to diffusive scattering. 

A  periodic ramp voltage ($0.1~\hertz, \pm 5V$ peak-to-peak) was applied to the Converter piezoelectric actuator, to encode a periodic phase-signature pattern among the outgoing time-bin states. The count pattern expected at the SPDA pixels in response to this phase-signature was pre-characterized as shown in \cref{fig:imaging_with_intensity}b. All the pixels followed roughly the same pattern. From now on, we shall refer to this as the \emph{reference pattern}. 

\Cref{fig:imaging_with_intensity}c shows the observed pattern at two neighbouring pixels $(4,3)$ and $(4,4)$. The observed patterns appear distorted due to low SNR, however the correlation with the reference pattern is clearly visible. A similar correlation was observed for all the pixels. However, since bright pixels (such as $(4,4)$) receive more photons than dark pixels (such as $(4,3)$) -- due to the feature of the object -- they are affected less by the noise, i.e., less distorted.   This particular feature has been exploited to enhance the contrast as described next.

In order to simulate a noisy low-light environment, the laser was highly attenuated and a  lamp was placed in front of the camera to create a high level of background signals. The scattering surface rotation was kept at $\phi = 40\degree$. The image captured by the imaging array is shown in \cref{fig:imaging_with_intensity}d. As expected, the presence of high background noise severely degraded the image contrast. The observed pattern in response to applied phase-signature at the same two neighbouring pixels $(4,3)$ and $(4,4)$ are also included in figure 4d. Due to the high background, the DC level of the observed patterns have shifted up considerably, but the correlation with the reference pattern is still visible. For each pixel, we calculated the correlation between the observed and reference pattern (see Methods for more details). We have chosen a threshold of $2.4$, and \cref{fig:imaging_with_intensity}e shows only those pixels having a correlation higher than this value. The result is a reconstruction of the object image with much-enhanced contrast. We note that during measuring the observed patterns, only photons from the SL and LS paths have been considered. In other words, from the three peaks shown in \cref{fig:spatial_temporal}, we only post-selected the detections from the middle-peak which required the timing data for each pixel. As a result, our approach is a combination of utilizing arrival times for coherence analysis, that can also be utilized for LIDAR applications.

We emphasize that instead of intensity, we used the correlation between observed and reference pattern to differentiate the bright pixels from the dark, i.e., instead of choosing an intensity threshold, we chose a correlation threshold. When intensity is used as a detection parameter, the contrast degrades with decreasing SNR. On the other hand, visibility -- or more specifically correlation -- does not drop as much with decreasing SNR as seen in \cref{fig:scattering_plot}b. Nevertheless, as a single parameter, visibility cannot generate the image as all the illuminated pixels roughly have the same visibility (see \cref{fig:patterns}a in Methods). But correlation drops with low SNR as pixels with low SNR are distorted more. As a result, the correlation is an efficient parameter -- combined with arrival time data  -- for imaging an object with enhanced contrast.

We would like to note that the above approach is just a proof-of-principle way to demonstrate how the quantum coherence observed from scattered photons could be used in enhancing the contrast.  It is in no way the optimum approach, and better approaches could certainly be devised. For example, the Analyzer interferometer could be realigned to access both the outputs having intensity, $P_{\pm}(\Theta) = S_{\pm}(\Theta) + N/2$, with $P_{\pm}(\Theta) (N/2)$ being the intensity (background noise) at the two output arms for a phase difference $\Theta$ (here, channel noise is assumed to be equally divided among the output arms). Then, instead of using statistics from a single-arm, measurements from both arms like $P(\Theta) = P_{+}(\Theta) - P_{-}(\Theta)$, could be used to nullify the effect of noise to improve the result. However, finding a more sophisticated image-reconstruction method is out of this paper's scope and will be analyzed elsewhere. 

\section*{Discussion}
\label{conclusion}
This work has demonstrated a novel and robust approach to transfer and recover quantum coherence via scattered photons by realizing a multimode, imaging time-bin interferometer equipped with a single-photon-detector-array sensor. Our quantum receiver achieves excellent temporal precision of $\approx 120~\pico\second$ as well as the ability to spatially resolve the field-of-view with excellent time-bin visibility across the whole sensor area. Each pixel  independently received the coherence from the spatial modes of the scattered photons. The maximum observed visibility was $95\%$, which remained within a $<10\%$ variation over a wide scattering angle range of $-45\degree$ to $+45\degree$. recovered from photons through a scattering, non-line-of-sight channel. All these features have the potential to open up new avenues in many applications, including quantum communication around-the-corner, low-light and 3D imaging, background noise rejection, around-the-corner imaging~\cite{gariepy2016}, velocity measurement~\cite{erksine1995}, LIDAR, object detection and identification, etc. 

We demonstrated the application potential of our method by showing two potential applications, one is non-line-of-sight quantum communications, the other is enhancing the contrast of single-photon images.  A more detailed and quantitative analysis of these applications will be given elsewhere. We believe our results will instigate further research on the application of coherence in quantum sensing, imaging, and communication and lead to novel areas of application. 

\section*{Author contribution}
S.S. performed the analyses and experiments. T.J. conceptualized the idea and supervised the study. Both authors contributed to writing the manuscript.

\section*{Acknowledgement}
We thank Duncan England, Bhashyam Balaji, Ramy Tannous, Youn Seok Lee
 and Alex Kirillova for discussions and technical support. This work was
supported by the National Research Council Canada, Defence Research Development Canada, Industry Canada, Canada Fund for Innovation, Ontario MRI, Ontario Research Fund, and NSERC (programs Discovery, CryptoWorks21, Strategic Partnership Grant) and Canada First Research Excellence Fund (TQT).

\section*{Disclosures}
The authors declare no conflicts of interest.

\def\bibsection{\medskip\begin{center}\rule{0.5\columnwidth}{.8pt}\end{center}\medskip} 

\begin{thebibliography}{10}
	\expandafter\ifx\csname url\endcsname\relax
	\def\url#1{\texttt{#1}}\fi
	\expandafter\ifx\csname urlprefix\endcsname\relax\def\urlprefix{URL }\fi
	\providecommand{\bibinfo}[2]{#2}
	\providecommand{\eprint}[2][]{\url{#2}}
	
	\bibitem{rubenok2013}
	\bibinfo{author}{Rubenok, A.}, \bibinfo{author}{Slater, J.~A.},
	\bibinfo{author}{Chan, P.}, \bibinfo{author}{Lucio-Martinez, I.} \&
	\bibinfo{author}{Tittel, W.}
	\newblock \bibinfo{title}{Real-world two-photon interference and
		proof-of-principle quantum key distribution immune to detector attacks}.
	\newblock \emph{\bibinfo{journal}{Phys. Rev. Lett.}}
	\textbf{\bibinfo{volume}{111}}, \bibinfo{pages}{130501}
	(\bibinfo{year}{2013}).
	\newblock
	\urlprefix\url{https://link.aps.org/doi/10.1103/PhysRevLett.111.130501}.
	
	\bibitem{michler2000}
	\bibinfo{author}{Michler, P.} \emph{et~al.}
	\newblock \bibinfo{title}{A quantum dot single-photon turnstile device}.
	\newblock \emph{\bibinfo{journal}{Science}} \textbf{\bibinfo{volume}{290}},
	\bibinfo{pages}{2282--2285} (\bibinfo{year}{2000}).
	\newblock \urlprefix\url{https://science.sciencemag.org/content/290/5500/2282}.
	\newblock
	\eprint{https://science.sciencemag.org/content/290/5500/2282.full.pdf}.
	
	\bibitem{lee2002}
	\bibinfo{author}{Lee, H.}, \bibinfo{author}{Kok, P.} \&
	\bibinfo{author}{Dowling, J.~P.}
	\newblock \bibinfo{title}{A quantum rosetta stone for interferometry}.
	\newblock \emph{\bibinfo{journal}{Journal of Modern Optics}}
	\textbf{\bibinfo{volume}{49}}, \bibinfo{pages}{2325--2338}
	(\bibinfo{year}{2002}).
	\newblock \urlprefix\url{https://doi.org/10.1080/0950034021000011536}.
	\newblock \eprint{https://doi.org/10.1080/0950034021000011536}.
	
	\bibitem{fulvio2015}
	\bibinfo{author}{Flamini, F.} \emph{et~al.}
	\newblock \bibinfo{title}{Thermally reconfigurable quantum photonic circuits at
		telecom wavelength by femtosecond laser micromachining}.
	\newblock \emph{\bibinfo{journal}{Light: Science \& Applications}}
	\textbf{\bibinfo{volume}{4}}, \bibinfo{pages}{e354--e354}
	(\bibinfo{year}{2015}).
	
	\bibitem{pirandola2015}
	\bibinfo{author}{Pirandola, S.}, \bibinfo{author}{Eisert, J.},
	\bibinfo{author}{Weedbrook, C.}, \bibinfo{author}{Furusawa, A.} \&
	\bibinfo{author}{Braunstein, S.~L.}
	\newblock \bibinfo{title}{Advances in quantum teleportation}.
	\newblock \emph{\bibinfo{journal}{Nature Photonics}}
	\textbf{\bibinfo{volume}{9}}, \bibinfo{pages}{641--652}
	(\bibinfo{year}{2015}).
	
	\bibitem{xu2015}
	\bibinfo{author}{Xu, F.} \emph{et~al.}
	\newblock \bibinfo{title}{Experimental quantum fingerprinting}
	\bibinfo{note}{{arXiv:1503.05499v1}}, \eprint{1503.05499v1}.
	
	\bibitem{irvine2004}
	\bibinfo{author}{Irvine, W. T.~M.}, \bibinfo{author}{Lamas~Linares, A.},
	\bibinfo{author}{de~Dood, M. J.~A.} \& \bibinfo{author}{Bouwmeester, D.}
	\newblock \bibinfo{title}{Optimal quantum cloning on a beam splitter}.
	\newblock \emph{\bibinfo{journal}{Phys. Rev. Lett.}}
	\textbf{\bibinfo{volume}{92}}, \bibinfo{pages}{047902}
	(\bibinfo{year}{2004}).
	\newblock
	\urlprefix\url{https://link.aps.org/doi/10.1103/PhysRevLett.92.047902}.
	
	\bibitem{pan2012}
	\bibinfo{author}{Pan, J.-W.} \emph{et~al.}
	\newblock \bibinfo{title}{Multiphoton entanglement and interferometry}.
	\newblock \emph{\bibinfo{journal}{Rev. Mod. Phys.}}
	\textbf{\bibinfo{volume}{84}}, \bibinfo{pages}{777--838}
	(\bibinfo{year}{2012}).
	\newblock \urlprefix\url{https://link.aps.org/doi/10.1103/RevModPhys.84.777}.
	
	\bibitem{spring2013}
	\bibinfo{author}{Spring, J.~B.} \emph{et~al.}
	\newblock \bibinfo{title}{Boson sampling on a photonic chip}.
	\newblock \emph{\bibinfo{journal}{Science}} \textbf{\bibinfo{volume}{339}},
	\bibinfo{pages}{798--801} (\bibinfo{year}{2013}).
	\newblock \urlprefix\url{https://science.sciencemag.org/content/339/6121/798}.
	
	\bibitem{gariepy2016}
	\bibinfo{author}{Gariepy, G.}, \bibinfo{author}{Tonolini, F.},
	\bibinfo{author}{Henderson, R.}, \bibinfo{author}{Leach, J.} \&
	\bibinfo{author}{Faccio, D.}
	\newblock \bibinfo{title}{Detection and tracking of moving objects hidden from
		view}.
	\newblock \emph{\bibinfo{journal}{Nature Photonics}}
	\textbf{\bibinfo{volume}{10}}, \bibinfo{pages}{23--26}
	(\bibinfo{year}{2016}).
	
	\bibitem{erksine1995}
	\bibinfo{author}{Erskine, D.~J.} \& \bibinfo{author}{Holmes, N.~C.}
	\newblock \bibinfo{title}{White-light velocimetry}.
	\newblock \emph{\bibinfo{journal}{Nature}} \textbf{\bibinfo{volume}{377}},
	\bibinfo{pages}{317--320} (\bibinfo{year}{1995}).
	
	\bibitem{Hohn:1969cr}
	\bibinfo{author}{H\"ohn, D.}
	\newblock \bibinfo{title}{Depolarization of a laser beam at 6328 a due to
		atmospheric transmission}.
	\newblock \emph{\bibinfo{journal}{Appl. Opt.}} \textbf{\bibinfo{volume}{8}},
	\bibinfo{pages}{367} (\bibinfo{year}{1969}).
	
	\bibitem{bourgoin2014}
	\bibinfo{author}{Bourgoin, J.-P.} \emph{et~al.}
	\newblock \bibinfo{title}{Experimentally simulating quantum key distribution
		with ground-to-satellite channel losses and processing limitations}
	\bibinfo{note}{({m}anuscript in preparation)}.
	
	\bibitem{sit2017}
	\bibinfo{author}{Sit, A.} \emph{et~al.}
	\newblock \bibinfo{title}{High-dimensional intracity quantum cryptography with
		structured photons}.
	\newblock \emph{\bibinfo{journal}{Optica}} \textbf{\bibinfo{volume}{4}},
	\bibinfo{pages}{1006--1010} (\bibinfo{year}{2017}).
	\newblock
	\urlprefix\url{http://www.osapublishing.org/optica/abstract.cfm?URI=optica-4-9-1006}.
	
	\bibitem{patterson2005}
	\bibinfo{author}{Paterson, C.}
	\newblock \bibinfo{title}{Atmospheric turbulence and orbital angular momentum
		of single photons for optical communication}.
	\newblock \emph{\bibinfo{journal}{Phys. Rev. Lett.}}
	\textbf{\bibinfo{volume}{94}}, \bibinfo{pages}{153901}
	(\bibinfo{year}{2005}).
	\newblock
	\urlprefix\url{https://link.aps.org/doi/10.1103/PhysRevLett.94.153901}.
	
	\bibitem{malik2012}
	\bibinfo{author}{Malik, M.} \emph{et~al.}
	\newblock \bibinfo{title}{Influence of atmospheric turbulence on optical
		communications using orbital angular momentum for encoding}.
	\newblock \emph{\bibinfo{journal}{Opt. Express}} \textbf{\bibinfo{volume}{20}},
	\bibinfo{pages}{13195--13200} (\bibinfo{year}{2012}).
	\newblock
	\urlprefix\url{http://www.opticsexpress.org/abstract.cfm?URI=oe-20-12-13195}.
	
	\bibitem{brendel1999}
	\bibinfo{author}{Brendel, J.}, \bibinfo{author}{Gisin, N.},
	\bibinfo{author}{Tittel, W.} \& \bibinfo{author}{Zbinden, H.}
	\newblock \bibinfo{title}{Pulsed energy-time entangled twin-photon source for
		quantum communication}.
	\newblock \emph{\bibinfo{journal}{Phys. Rev. Lett.}}
	\textbf{\bibinfo{volume}{82}}, \bibinfo{pages}{2594--2597}
	(\bibinfo{year}{1999}).
	\newblock \urlprefix\url{https://link.aps.org/doi/10.1103/PhysRevLett.82.2594}.
	
	\bibitem{stucki2005}
	\bibinfo{author}{Stucki, D.}, \bibinfo{author}{Brunner, N.},
	\bibinfo{author}{Gisin, N.}, \bibinfo{author}{Scarani, V.} \&
	\bibinfo{author}{Zbinden, H.}
	\newblock \bibinfo{title}{Fast and simple one-way quantum key distribution}.
	\newblock \emph{\bibinfo{journal}{Appl. Phys. Lett.}}
	\textbf{\bibinfo{volume}{87}}, \bibinfo{pages}{194108}
	(\bibinfo{year}{2005}).
	
	\bibitem{townsend1993}
	\bibinfo{author}{Townsend, P.~D.}, \bibinfo{author}{Rarity, J.~G.} \&
	\bibinfo{author}{Tapster, P.~R.}
	\newblock \bibinfo{title}{Single photon interference in 10 km long optical
		fibre interferometer}.
	\newblock \emph{\bibinfo{journal}{Electron. Lett.}}
	\textbf{\bibinfo{volume}{29}}, \bibinfo{pages}{634--635}
	(\bibinfo{year}{1993}).
	
	\bibitem{muller1997}
	\bibinfo{author}{Muller, A.} \emph{et~al.}
	\newblock \bibinfo{title}{{``Plug and play''} systems for quantum
		cryptography}.
	\newblock \emph{\bibinfo{journal}{Appl. Phys. Lett.}}
	\textbf{\bibinfo{volume}{70}}, \bibinfo{pages}{793--795}
	(\bibinfo{year}{1997}).
	
	\bibitem{inoue2002}
	\bibinfo{author}{Inoue, K.}, \bibinfo{author}{Waks, E.} \&
	\bibinfo{author}{Yamamoto, Y.}
	\newblock \bibinfo{title}{Differential phase shift quantum key distribution}.
	\newblock \emph{\bibinfo{journal}{Phys. Rev. Lett.}}
	\textbf{\bibinfo{volume}{89}}, \bibinfo{pages}{037902}
	(\bibinfo{year}{2002}).
	
	\bibitem{stucki2009}
	\bibinfo{author}{Stucki, D.} \emph{et~al.}
	\newblock \bibinfo{title}{High rate, long-distance quantum key distribution
		over 250 km of ultra low loss fibres}.
	\newblock \emph{\bibinfo{journal}{New J. Phys.}} \textbf{\bibinfo{volume}{11}},
	\bibinfo{pages}{075003} (\bibinfo{year}{2009}).
	
	\bibitem{jin2018}
	\bibinfo{author}{Jin, J.} \emph{et~al.}
	\newblock \bibinfo{title}{Demonstration of analyzers for multimode photonic
		time-bin qubits}.
	\newblock \emph{\bibinfo{journal}{Phys. Rev. A}} \textbf{\bibinfo{volume}{97}},
	\bibinfo{pages}{043847} (\bibinfo{year}{2018}).
	\newblock \urlprefix\url{https://link.aps.org/doi/10.1103/PhysRevA.97.043847}.
	
	\bibitem{jin2019}
	\bibinfo{author}{Jin, J.} \emph{et~al.}
	\newblock \bibinfo{title}{Genuine time-bin-encoded quantum key distribution
		over a turbulent depolarizing free-space channel}.
	\newblock \emph{\bibinfo{journal}{Opt. Express}} \textbf{\bibinfo{volume}{27}},
	\bibinfo{pages}{37214 -- 37223} (\bibinfo{year}{2019}).
	\newblock
	\urlprefix\url{http://www.opticsexpress.org/abstract.cfm?URI=oe-27-26-37214}.
	
	\bibitem{vallone2016}
	\bibinfo{author}{Vallone, G.} \emph{et~al.}
	\newblock \bibinfo{title}{Interference at the single photon level along
		satellite-ground channels}.
	\newblock \emph{\bibinfo{journal}{Phys. Rev. Lett.}}
	\textbf{\bibinfo{volume}{116}}, \bibinfo{pages}{253601}
	(\bibinfo{year}{2016}).
	\newblock
	\urlprefix\url{https://link.aps.org/doi/10.1103/PhysRevLett.116.253601}.
	
	\bibitem{cao2020}
	\bibinfo{author}{Cao, Y.} \emph{et~al.}
	\newblock \bibinfo{title}{Long-distance free-space
		measurement-device-independent quantum key distribution} \eprint{2006.05088}.
	
	\bibitem{tan2017}
	\bibinfo{author}{Tan, K.} \& \bibinfo{author}{Cheng, X.}
	\newblock \bibinfo{title}{Specular reflection effects elimination in
		terrestrial laser scanning intensity data using phong model}.
	\newblock \emph{\bibinfo{journal}{Remote Sensing}} \textbf{\bibinfo{volume}{9}}
	(\bibinfo{year}{2017}).
	\newblock \urlprefix\url{https://www.mdpi.com/2072-4292/9/8/853}.
	
	\bibitem{bennett1984}
	\bibinfo{author}{Bennett, C.~H.} \& \bibinfo{author}{Brassard, G.}
	\newblock \bibinfo{title}{Quantum cryptography: Public key distribution and
		coin tossing}.
	\newblock In \emph{\bibinfo{booktitle}{Proceedings of IEEE International
			Conference on Computers, Systems, and Signal Processing}},
	\bibinfo{pages}{175--179} (\bibinfo{publisher}{IEEE Press, New York},
	\bibinfo{address}{Bangalore, India}, \bibinfo{year}{1984}).
	
	\bibitem{lydersen2010a}
	\bibinfo{author}{Lydersen, L.} \emph{et~al.}
	\newblock \bibinfo{title}{Hacking commercial quantum cryptography systems by
		tailored bright illumination}.
	\newblock \emph{\bibinfo{journal}{Nat. Photonics}}
	\textbf{\bibinfo{volume}{4}}, \bibinfo{pages}{686--689}
	(\bibinfo{year}{2010}).
	
	\bibitem{lydersen2011}
	\bibinfo{author}{Lydersen, L.}, \bibinfo{author}{Skaar, J.} \&
	\bibinfo{author}{Makarov, V.}
	\newblock \bibinfo{title}{Tailored bright illumination attack on
		distributed-phase-reference protocols}.
	\newblock \emph{\bibinfo{journal}{J. Mod. Opt.}} \textbf{\bibinfo{volume}{58}},
	\bibinfo{pages}{680--685} (\bibinfo{year}{2011}).
	
	\bibitem{gras2020}
	\bibinfo{author}{Gras, G.}, \bibinfo{author}{Rusca, D.},
	\bibinfo{author}{Zbinden, H.} \& \bibinfo{author}{Bussieres, F.}
	\newblock \bibinfo{title}{Countermeasure against quantum hacking using
		detection statistics} \eprint{2010.08474}.
	
	\bibitem{sajeed2015a}
	\bibinfo{author}{Sajeed, S.} \emph{et~al.}
	\newblock \bibinfo{title}{Security loophole in free-space quantum key
		distribution due to spatial-mode detector-efficiency mismatch}.
	\newblock \emph{\bibinfo{journal}{Phys. Rev. A}} \textbf{\bibinfo{volume}{91}},
	\bibinfo{pages}{062301} (\bibinfo{year}{2015}).
	
	\bibitem{sascha2017}
	\bibinfo{author}{{Agne, Sascha}}.
	\newblock \emph{\bibinfo{title}{Exploration of Higher-Order Quantum
			Interference Landscapes}}.
	\newblock Ph.D. thesis (\bibinfo{year}{2017}).
	\newblock \urlprefix\url{http://hdl.handle.net/10012/12307}.
	
\end{thebibliography}

%
\section*{Methods}

\subsection*{Detailed experimental setup}
The experimental setup consists of two unbalanced Michelson interferometers (UMIs). The first UMI -- \emph{Converter} -- creates the coherence while the second one  -- \emph{Analyzer} -- measures the coherence. The laser (Picoquant laser diode LDH 8-1-596, PDL-800B driver) emits $697~\nano\meter$, $300\pico\second$ multimode light pulses at a rate of $5$MHz. Lights coming out of the multimode fiber are collimated and sent through the Converter. The path difference $\Delta_C$ between the long and short arms turns each incoming state into a coherent superposition of two time-bins separated by $0.57 \nano \second$. They are sent towards the scattering surface and the scattered photons are collected by the Analyzer. The path difference between the short and the long arm in the Analyzer is $\Delta_A$. A photon coming out of the Analyzer could have traversed one of the four possible paths: short-short (SS), short-long (SL), long-short (LS) and long-long (LL). If the path-differences in each UMI match, i.e., $\Delta_C \approx \Delta_A$, the probabilities of a photon coming via SL and LS paths become indistinguishable and result in interference. Piezoelectric crystals are placed in the short arm of each interferometer to adjust the path difference. The photons emerging from the Analyzer are focused into an SPDA containing $64$ single-photon avalanche photo-diodes. The pixels of the SPDA are arranged in a $8\times 8$ row-column configuration, with a pitch of $75~\micro\meter$. A focusing lens (Canon EF-S 18-200mm f/3.5-5.6) is used to illuminate the desired range of pixels with its focal length set to about 60~mm, thus yielding an angular resolution of $0.07~\degree$, and a total angular field of $0.5~\degree$. 


%
\begin{figure}[htp]
	\centering
	\includegraphics[width=.95\columnwidth]{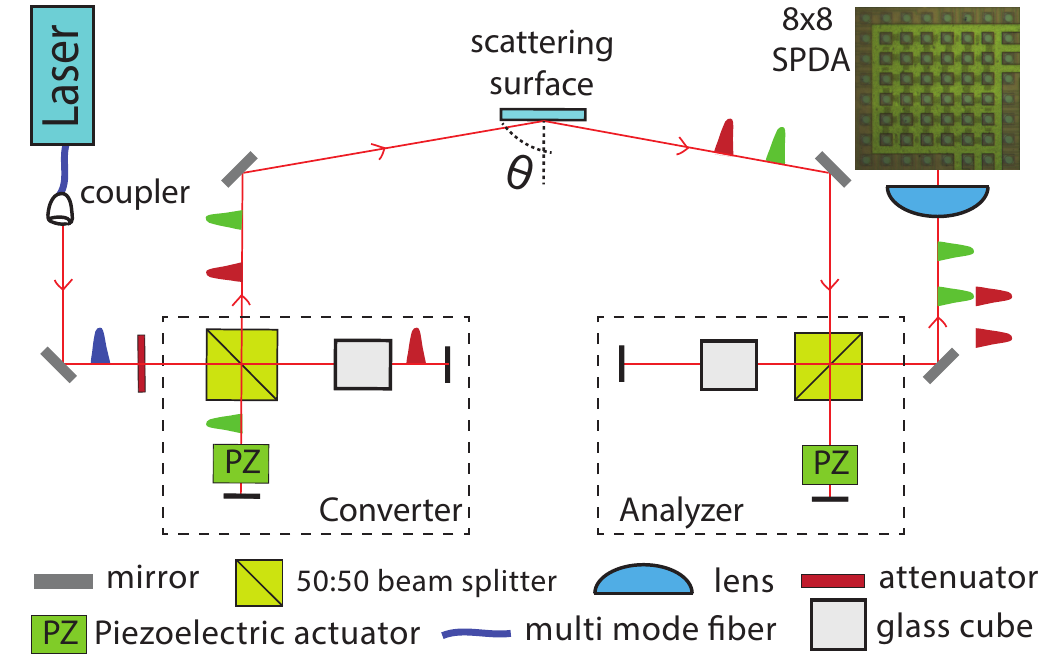}
	\caption{The experimental setup. The Converter creates two pulses seperated by $0.6\nano\second$. The path difference between the two short-long (SL) and long-short (LS) pulses can be adjusted using the piezoelectric actuator placed in the short arm. A glass cube is present in the long arm of both interferometers to make the interferometers balanced in the spatial domain. The Analyzer recombines the pulses coming via SL and LS paths and the output is studied by the $8 \times 8$ SPDA. During the initial alignment, the incident angle $\theta = 25~\degree$ (figure not to scale). }
	\label{fig:setup_method} 
\end{figure}

\subsection*{Compensation for spatial mode distortion}
\label{balancing}
A time-bin encoded photon entering an Analyzer with variable angle of incidence causes a lateral offset between the paths impinging at the exit beam-splitter~\cite{sascha2017,jin2018} causing degradation of interference visibility. Channel induced spatial-mode distortions further lower the interference quality. The root cause of this is the inherent asymmetry of the unbalanced Michelson interferometers. A compensation technique is to use a glass material with appropriate length and refractive index to create a virtual mirror closer to the beamsplitter. In this way, the interferometer, although asymmetric in time, becomes symmetric in the spatial-mode domain.  We placed a 118 mm-long glass cube with refractive index 1.4825 at the long arm to match the distance between beam-splitter-to-virtual-mirror with the short arm (see \cite{sascha2017} for more details and the analytical formula). This compensation not only improves performance at higher AOI, but is also necessary to enable high interference visibility with a multimode beam.

\subsection*{Correlation of patterns}
\label{pattern_correlation}
\begin{figure}[tp]
	\centering
	\includegraphics[width=\columnwidth]{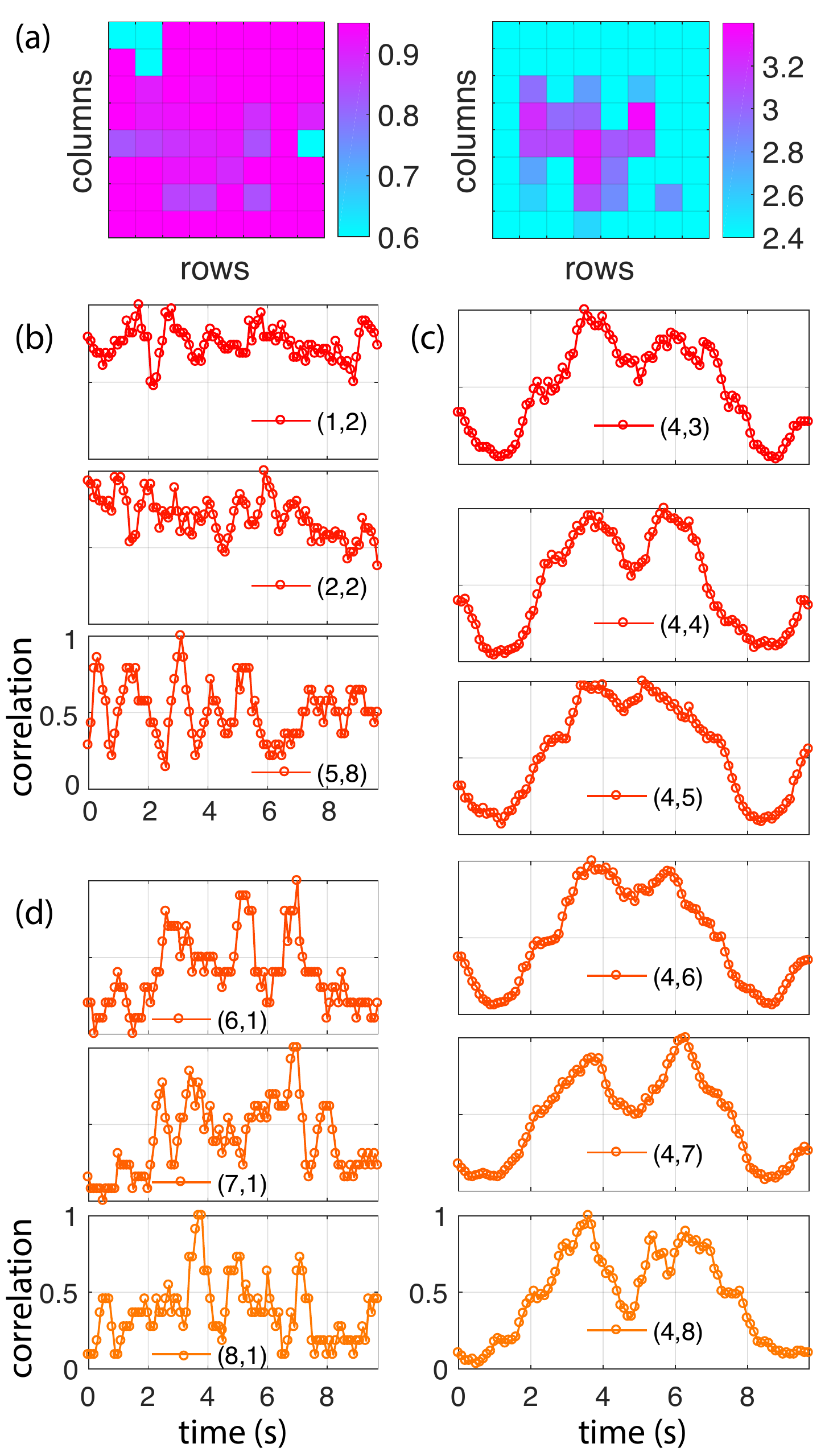}
	\caption{Visibility and patterns observed at different pixels of the SPDA imager. a. visibility across all the pixels (left) and the reconstructed image drawn from correlation (right) b. observed pattern at the three defective pixels having high dark counts. The correlation with the reference pattern is lower than threshold. c. observed pattern at the bottom six pixels of column four. The correlation with the reference pattern is higher than the threshold. d. observed pattern at bottom three pixels of column one. The correlation with the reference pattern is between the two previous extremes but still lower than the chosen threshold. By plotting the pixels with correlation higher than the threshold, the image was reconstructed with enhanced contrast as shown in fig a (right).}
	\label{fig:patterns}.
\end{figure}

Here we describe the method by which \cref{fig:imaging_with_intensity}e was plotted.  \Cref{fig:patterns}a shows the same image (right) along with the observed visibility (left) across all the SPDA pixels. The visibility was calculated as $ V = (I_{max} - i_{min}) / (I_{max} + I_{min})$ where $I_{max}~(I_{min})$ is the maximum (minimum) count in a phase-signature cycle. Pixel $(1,1)$ was the trigger, and $(1,2), (2,2)$ and $(5,8)$ had very high dark counts.  The observed pattern at these three defected pixels are shown in \Cref{fig:patterns}b where no correlation with the reference pattern was visible. Fig. c shows the observed pattern at the bottom six pixels of column 4. The resemblance with the \emph{reference pattern} was quite pronounced. On the other hand, fig d shows the observed patterns for the bottom three pixels of column 1 that detected relatively fewer number of photons and are much more distorted. They are in the middle of the two previous extremes and the correlation value with the reference pattern for them lies in the middle. Before calculating the cross-correlation, both the reference and observed patterns were shifted downward by $0.5$ for the ease of analysis.

\subsection*{The single-photon-detector-array sensor}
\label{spda}
\begin{figure*}[htp]
	\centering
	\includegraphics[width=.8\textwidth]{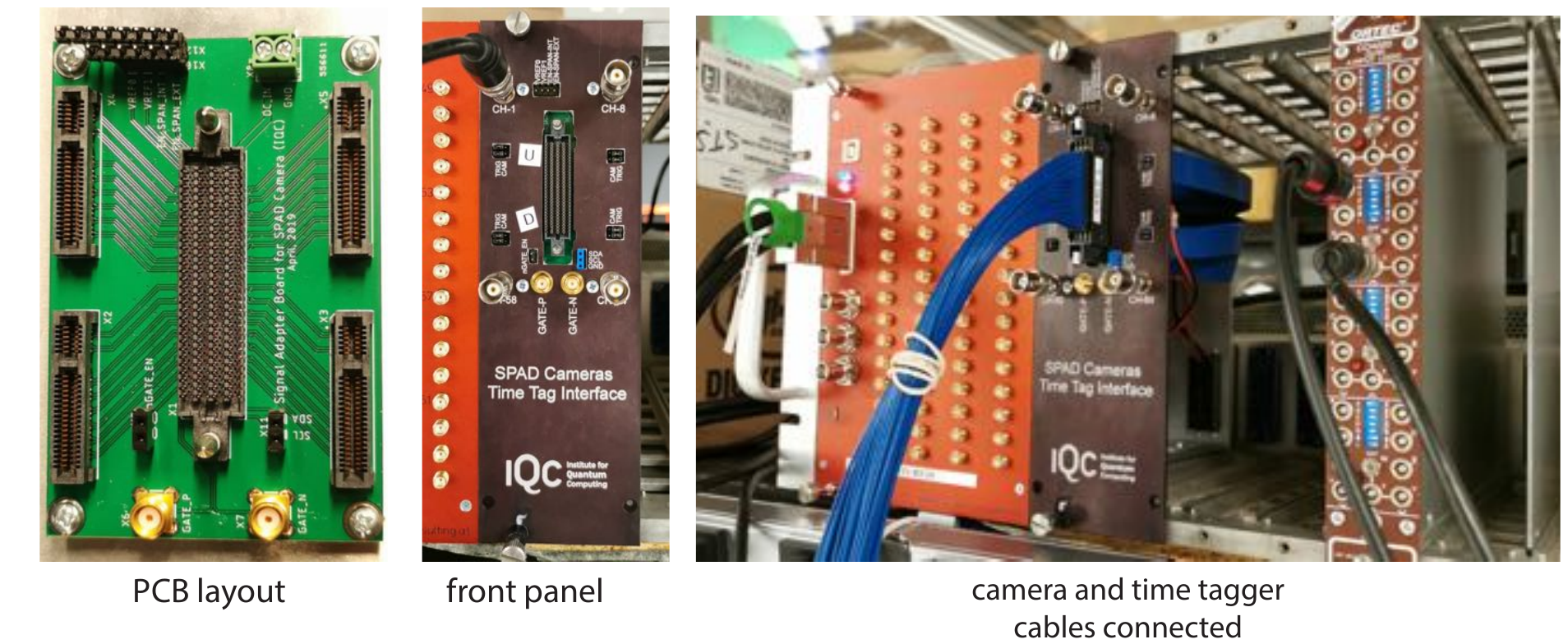}	
	\caption{Interface board. The camera output (64 LVDS pairs) are linked to an interface board, and divides the signals to 4 blocks of 16 LVDS pairs, as well as provides control signal inputs. The 4 signal blocks are directly interfaced with a high-speed time-tagger that can handle 64 input channels, up to 1 GTag/s.}
	\label{interface-board}
\end{figure*}

The SPDA camera is composed of 64 single photon avalanche diodes arranged in a square pattern with 8 rows and 8 columns. Each row and column has eight $30\micro\meter$ diameter detectors with X and Y pitch equal to $75~\micro\meter$. The average dark count rate was $35$~cps, deadtime was $150~\nano\second$ and the timing resolution was $100\pico\second$. The electrical output from the camera (Samtec Cable, SEAC-30-08-XX.X-TU-TU-2) contains 64 LVDS detection signal from the pixels along with other control signals. An adapter board was designed to match the electrical signal interface of the camera to that of the time tagger (Universal quantum devices) as shown in \cref{interface-board}. The developed board translates the $64$ LVDS camera outputs into four groups of $16$ LVDS channels going into the time-tagger.  As a result, time stamps from the camera can be read by the tagger at a rate of approximately $1$ GTag/s through an external-PCI-express interface. The adapter board provide the option to sacrifice upto four pixels ($1, 8, 58 $ and $64$) to be used as external trigger. In this work only one trigger was used. When triggering is not required, the pixels can function as normal detectors. 

\end{document}